\documentstyle[12pt,aaspp]{article}
\eqsecnum
\journalid{}{}
\articleid{}{}
\slugcomment{Accepted for {\it The Astronomical Journal}, $\approx$ July, 1996}

\def\vi{\hbox{$V\!-\!I$}} 
\def\vk{\hbox{$V\!-\!K$}} 
\def\jk{\hbox{$J\!-\!K$}} 

\def\mfe{$\langle{\rm Fe}\rangle$}

\def\teff{$T_{\rm e}$}
\def\gg{log $g$}
\def\feh{[Fe/H]}
\def\evi{E$_{V-I}$}

\def\etal{{\it et al.} }

\def\lta{\mathrel{\spose{\lower 3pt\hbox{$\mathchar"218$}}
     \raise 2.0pt\hbox{$\mathchar"13C$}}}
\def\gta{\mathrel{\spose{\lower 3pt\hbox{$\mathchar"218$}}
     \raise 2.0pt\hbox{$\mathchar"13E$}}}

\begin{document}

\title{K Giants in Baade's Window. II.  The Abundance
Distribution\footnote{Based on data obtained at the Anglo--Australian
Observatory and at the Cerro Tololo Inter--American Observatory, NOAO,
which is operated by the Association of Universities for Research in
Astronomy, Inc. (AURA), under cooperative agreement with the National
Science Foundation.}
}

\author{Elaine M. Sadler}
\affil{School of Physics, University of Sydney,
NSW 2006, Australia \\
Electronic mail: ems@physics.usyd.edu.au}

\author{R. Michael Rich\altaffilmark{2}}
\affil{Department of Astronomy, Pupin Laboratories, Columbia University, \\
538 West 120th Street, New York, New York  10027 \\
Electronic mail: rmr@cuphyd.columbia.edu}

\author{D. M. Terndrup\altaffilmark{3}}
\affil{Department of Astronomy, The Ohio State University, \\
174 W. 18th Ave., Columbus, Ohio 43210 \\
Electronic mail: terndrup@baade.mps.ohio--state.edu}

\altaffiltext{2}{Alfred P. Sloan Fellow.}
\altaffiltext{3}{Presidential Young Investigator.}

\begin{abstract}

This is the second in a series of papers in which we analyze spectra of
over 400 K and M giants in Baade's Window, including most of the stars
with proper motions measured by Spaenhauer {\it et al.} [AJ, 103, 297
(1992)].  In our first paper, we measured line--strength indices of Fe,
Mg, CN and H$\beta$ and calibrated them on the system of Faber {\it et
al.} [ApJS, 57, 711 (1985)].  Here, we use the $\langle{\rm Fe}\rangle$
index to derive an abundance distribution in [Fe/H] for 322 stars with
effective temperatures between 3900 K and 5160 K.

Our derived values of \feh\ agree well with those measured from
high--resolution echelle spectra (e.g., McWilliam \& Rich [ApJS, 91,
749 (1994)]) for the small number of stars in common.  We find a mean
abundance $\langle{\rm [Fe/H]}\rangle = -0.11 \pm 0.04$ for our sample
of Baade's Window K giants. More than half the sample lie in the range
$-0.4 <$ \feh\ $<+0.3$.

We estimate line--of--sight distances for individual stars in our
sample and confirm that, in Baade's Window, most K giants with $V <
15.5$ are foreground disk stars, but the great majority (more than
80\%) with $V > 16$ belong to the bulge.

We also compare the metallicities derived from the CN and Mg$_2$
indices to those from iron.  Most of the metal--rich stars in our
sample appear to be CN--weak, in contrast to the situation in
metal--rich globular clusters and elliptical galaxies.  The metal--poor
half of our sample ([Fe/H] $< 0$) shows evidence for a mild Mg
overenhancement ([Mg/Fe] $\sim +0.2$); but this is not seen in the more
metal--rich stars ([Fe/H] $\geq$ 0).  The K giants in Baade's Window
therefore share some, but not all, of the characteristics of stars in
elliptical galaxies as inferred from their integrated light.

\end{abstract} 


\section{Introduction}

This is the second paper of a series on the radial velocities and
metallicities of a large sample of K giants in the Baade's Window (BW)
field of the Galactic nuclear bulge at $(\ell, b) = (1^\circ,
-3.9^\circ)$.  In the first paper of this series (\cite{ter95},
hereafter Paper I), we presented line--strength, radial velocity and
photometric measurements for over 400 K and M--giant stars in BW for
which proper motions have already been measured (\cite{spa92}), and
placed the indices on the Lick Observatory system (\cite{fab85},
hereafter FFBG).  The main goals of this paper are to derive the
abundance distribution for the K--giant stars in our sample, and to
examine the distribution of the stars in distance along the line of
sight.  Our main abundance indicator is the iron--line index \mfe.  We
use \vi\ colors (corrected for extinction) to derive the effective
temperature \teff\ for each star, then apply the empirical calibration
derived by FFBG to determine \feh\ from the measured \mfe.  Subsequent
papers will discuss the three--dimensional kinematics of the bulge and
foreground disk in BW, and present the distribution of metallicities
and radial velocities along other lines of sight to the bulge.

The paper is organized as follows: we first (Sec.\ 2) derive effective
temperatures, then determine \feh\ for individual stars, and finally
compare our adopted metallicity scale to those from several previous
studies in BW.  We then estimate the distances of individual stars
(Sec.\ 3) and identify the main population subgroups in the sample.  In
Sec.\ 4, we address the question of non--solar abundance ratios in BW
stars via the Mg and CN indices.  We summarize and discuss the
implications of our results in Sec.\ 5.

\section{The Adopted Metallicity Scale}

\subsection{Philosophy}

In Paper I, we tabulated measurements of several of the line--strength
indices defined by FFBG and confirmed that our measurements are on the
same index system as defined by the standards in that paper.  We now
need to convert these indices into estimates of [Fe/H].  We adopt a
simple transformation based on the FFBG paper which relies on the
$\langle{\rm Fe}\rangle$ index, where $\langle{\rm Fe}\rangle = 0.5
\times ({\rm Fe~5270} + {\rm Fe~5335})$, and use a temperature scale
given by \vk.  We later demonstrate that this metallicity scale is in
agreement with other estimates in BW, particularly from high-resolution
spectra.  Alternative approaches for deriving [Fe/H] from
low-resolution spectra are discussed, for example, by \cite{jon95}.

Our choice of the FFBG metallicity calibration (which is separate from
the calibration of our line-strength {\it indices} on the FFBG system)
presents several problems.  The metallicity calibration, for example,
is given by the {\it adopted} values of [Fe/H] for a large number of
field stars, which are based on a sample of standard stars analyzed at
various resolutions over many years.  This calibration fundamentally
rests in turn on analyses of high--resolution spectra of a few of their
standard stars.  The high-resolution analyses available to FFBG
typically did not extend beyond [Fe/H] $> 0$; the few stars at very
high abundances typically had estimates based on low-resolution spectra
or narrow-band photometry which centered on one or two spectral
features, particularly CN.  Many of these analyses were done 30--40
years ago, and there has been considerable progress since then in the
study of field-star abundances.  Similarly at low-metallicities, the
line-strength indices become very weak and the derived metallicities
have relatively high errors.  It is therefore possible that future work
may revise the metallicity scale of the Lick indices and thus our
metallicity calibration for our sample.

Another limitation of the FFBG system is that, even with subsequent
extensions and modifications, there are indications that their
standards to not cover the full range of temperature, metallicity,
gravity, and non-solar abundance ratios which may be present throughout
the Galaxy.  The original FFBG calibration was extended significantly
by Gorgas {\it et al.} (1993, hereafter G93), who increased the number
of standards in the system and placed special emphasis on exploring
metallicity and gravity effects.  They computed polynomial ``fitting
functions'' which described the behavior of the FFBG indices as a
function of temperature\footnote{Worthey {\it et al.} (1994) presented
a modification to the G93 approach, in which the effective temperature
itself (specifically $\Theta_e = 5040 / T_{\rm e}$) was used as the
temperature parameter in fitting functions, rather than $V - K$.}
(parameterized by $V - K$), gravity ($\log g$), and metal abundance ($Z
\equiv {\rm [Fe/H]})$.  They showed that the Fe\,5270 and Fe\,5335
indices were sensitive to abundance but insensitive to gravity, while
the Mg indices were highly sensitive to both metallicity and gravity.
G93 described a procedure for deriving $Z$ and $\log g$ for individual
stars by averaging estimates from the CN, Mg$_2$ Fe\,5270, and Fe\,5335
indices.  For data with good signal--to--noise, this is accurate to
about 0.25 dex in $Z$ and 0.23 dex in $\log g$.   Despite increasing
evidence for non-solar abundance ratios in different galactic
populations, G93 concluded that only one metallicity parameter was
needed to characterize abundance variations among the different
populations of stars in the library.  They reached this conclusion by
demonstrating that the scatter about the (three parameter) fitting
functions was only slightly larger than the errors of measurement,
i.e., that there was no remaining variation which required the
introduction of other parameters, particularly those relating to
non-solar abundances ratios such as [Mg/Fe].

We decided to adopt the simpler FFBG approach, which relies primarily
on the \mfe\ index, for several reasons.  The first reason was that the
FFBG calibrations use simple linear or quadratic equations, which are
easier to compute than the more complicated fitting functions in G93.
The \mfe\ index itself, furthermore, is insensitive to gravity (G93),
which means that it is not necessary to know the distances to the
individual stars for an accurate estimate of [Fe/H].  In addition, we
presented evidence in Paper I that many stars in Baade's Window have
stronger Mg$_2$ absorption relative to \mfe\ than stars of the same
color in the G93 libraries.  We interpreted this to mean that if the
Baade's Window giants were relatively old ($t > 8 \times 10^9$ yr),
they have an enhanced relative abundance in Mg (i.e., [Mg/Fe] $> 0$).
In this case, an abundance measurement using a combination of Mg and Fe
lines would yield a higher metallicity than one which used Fe alone
(cf. \cite{mcw94}).  Equivalently, if the metallicity scale from the
iron and magnesium lines were equal (as in G93), then for moderate
distances from the Sun ($R \sim 8$ kpc) they would be more massive than
is allowed from studies of the main--sequence turnoff.  Since one of
the aims of this paper is to explore the possibility of non--solar
abundance ratios (below, Sec.\ 4), we wished to avoid the implied
equality between the Mg and Fe scales in G93.  Finally, the errors in
our measured values of \mfe\ for individual stars are often too large
for us to set effective constraints on the two free parameters ([Fe/H]
and \gg) in the G93 formalism.

We observed more than twenty of the abundance standard stars listed by
FFBG, but (as described in Paper I) these observations were used only
to transform our line--strength measurements of the BW stars to the
Lick system.  We were able to place our index values accurately on the
FFBG system.  For the eight standards (excluding HR 5270) observed at
the AAT, the mean difference between our derived abundance \feh\ and
the FFBG value was (AAT $-$ FFBG) $= -0.02$ dex, with a dispersion of
0.27 dex.  For the much larger sample of standards observed at CTIO,
the difference (CTIO $-$ FFBG) was $+0.06$ dex, with a dispersion of
0.25 dex. (The ${\rm Fe}~5270$ and ${\rm Fe}~5335$ indices were
transformed to the FFBG system separately, then averaged.)

\subsection{The Temperature Scale}

The first step deriving metallicities for the BW sample is to transform
each \vi\ color to an effective temperature \teff.  Several authors
have investigated the metallicity and gravity dependence of the
transformation between broad--band colors and \teff, and have generally
found similar relations between effective temperature and $V - I$ agree
well (e.g., \cite{wor94}, their Figure 2) at least for $T_e > 3800$ K.
For this study, we used an empirical temperature calibration of the
Cousins $VRI$ system (\cite{bes79}), and calculated \vi\ colors for a
grid of models of known temperature (\cite{bel89}).  These two scales
are similar, though the models are 0.08 mag bluer than Bessell's values
for solar-metallicity giants at the cool end of the K giant temperature
distribution (\teff $\approx 4000$ K).  Here we adopt the Bessell
relation between the reddening--corrected color (\vi)$_0$\ and \teff,
which is displayed in Figure 1.

We restrict our derivation of effective temperatures to the K stars in
our sample, because the relation between \vi\ and \teff\ is essentially
independent of surface gravity for K stars (\cite{bes79}), but not for
the cooler M stars.  The relation between temperature and broad--band
colors, as derived from model fluxes, is also largely independent of
metallicity for $T_{\rm e} > 3800$ K (e.g., \cite{bes89};
\cite{wor94} and references therein).  For the cooler M giants the
temperature scale is increasingly poorly defined, because of TiO
blanketing in the $V$ band and a lack of calibration stars with
reliably--determined temperatures (e.g., \cite{rid80}).  We therefore
chose to restrict our analysis to stars with \teff\ in the range
3900--5160\,K.

Aside from random errors in the photometry, there are two sources of
error in the \vi\ colors and the determination of effective
temperatures.  These were extensively discussed in Paper I.  The first
systematic error arises from uncertainties in the zero points of the
photometry on the several CCD frames from which colors were measured.
This error is of order $\approx 0.04$ mag on each frame.  The second
error comes from the uncertainty in the adopted reddening.  In paper I,
we derived an average reddening of \evi\ $= 0.64 \pm 0.08$ mag for the
entire BW sample, and showed that there is marginal evidence in our
data for variations in the reddening across the field.  The BW field
has large (and patchy) reddening (e.g., Figure 1 of \cite{bla84}); so
the average extinction value may not be appropriate for the individual
stars in our sample.  If we have overestimated the reddening to an
individual star then we will derive a temperature which is too hot and
an abundance which is too high.  Similarly, underestimating the
reddening will yield a temperature which is too cool and an abundance
estimate which is too low.  We discuss the effects of these
uncertainties at greater length below.

\subsection{Computation of [Fe/H]}

The next step in computing metallicities was to recast the FFBG
empirical calibration of the \mfe\ index, which represents temperature
by \vk\ color, into expressions in \teff.  We converted the \vk\ colors
listed by FFBG to \teff\ using stars whose effective temperature is
tabulated by Bell \& Gustafsson (1989). This allowed us to rewrite the
FFBG calibration in terms of \teff\ rather than \vk, and hence to apply
it to the BW stars since we have no \vk\ photometry for most of the
sample.\footnote{Alternatively, we could have used the empirically
derived relation between $V - I$ and $V - K$ for BW stars (Tiede
{\it et al.} 1995; Paper I) and used the FFBG calibration directly.
Experiments showed that this approach yielded nearly identical
results to the one we adopted here.}

For reasons discussed above, we then restricted our computation of
[Fe/H] to those stars with \teff\ between 3900 K and 5160 K,
corresponding to spectral types between K5 and G8.  (The FFBG
calibration is defined for a slightly wider temperature range, roughly
3800--5200 K.)  Of the 432 stars listed in Paper I, 369 have
\vi\ colors which place them in our restricted temperature range.  Of
the remaining 63 stars, 36 are cooler (M stars), 16 are hotter (G
stars), and we have no photometry for 11.  We were able to measure
\mfe\ for 322 of these 369 stars.  For the remainder, we either have no
spectrum or the \mfe\ index was unmeasurable because of a poor
signal--to--noise ratio in that region of the spectrum or (for the CTIO
spectra) a cosmic--ray hit.

For these 322 stars, we used \teff\ and the iron line strength \mfe\ to
derive a metal abundance [Fe/H] using our recast FFBG formulation.
FFBG's Table 4 gives the relation between \vk, [Fe/H] and a parameter
$\langle\Delta{\rm Fe}\rangle$, defined as the difference between the
observed $\langle{\rm Fe}\rangle$ and the value expected for a star of
solar abundance at the same temperature.  In Figure 2 we plot this in a
slightly different form, showing \mfe\ versus \teff\ for a range of
values of [Fe/H], along with the observed values.

Table 1 lists all the BW stars in our sample for which we were able to
measure \feh.  The errors in \feh\ were initially estimated from the
errors in the individual values of \mfe\ listed in Paper I.  However, a
cross--comparison of [Fe/H] for stars in common with previous bulge
studies (immediately below) suggested that our errors were
overestimated by almost a factor of two.  We have therefore adjusted
our original estimates to give the final errors quoted in Table 1.

The FFBG relations are only valid for stars with \feh\ from $-0.8$ to
$\sim +0.5$.   The FFBG system loses sensitivity at low metallicities
because the Fe features become weak; while the high abundances may be
in error because they are derived by extrapolating past the calibrated
metallicities (this is a separate issue from the {\it validity} of the
adopted metallicities in the FFBG system at high [Fe/H]).  About 7\% of
the stars in our sample have \feh\ below this range, and 14\% have
abundances above.  For these stars, the derived \feh\ values are likely
to be indicative only.

\subsection{Error Estimates and Comparison with Other Studies}
 
We have a number of stars in common with other studies of the
metallicity distribution in BW.  Figure 3 shows the correlation between
our [Fe/H] estimates and those from (top panel) Washington photometry
by Geisler \& Friel (1992; hereafter GF92) and (lower panel)
low--resolution spectra by Rich (1988, hereafter R88).   Our [Fe/H]
measurements are systematically lower than those from R88 and GF92 by
0.19 and 0.15 dex respectively (see Table 2 for a summary).  The
difference appears to be a simple zero--point shift, independent of
temperature and abundance (Figure 3).  Table 3 gives a
cross--identification for the GF92 stars in common with our sample.

Recently, McWilliam \& Rich (1994, hereafter MR94) have presented a
detailed analysis of high--resolution spectra of 11 K--giants in BW,
six of them in common with our sample.  Consistent with our findings
above, they showed that their metallicity scale was lower on average
than that of R88 and GF92 by $\approx 0.3$ dex.  They discuss the
reasons for this offset in some detail, and conclude that the most
likely explanation for the previous higher metallicity scale is that
the earlier work was measuring something more like [(Fe+Mg)/H], which
would be higher than [Fe/H] if [Mg/Fe] $> 0$.  Denoting our
metallicities from Table 1 by SRT, we find that $\langle$\feh$_{\rm
SRT} - $\feh$_{\rm MR94}\rangle = +0.02 \pm 0.29$ (s.d.), showing that
the simple FFBG metallicity estimates we adopt here agree well with
those derived by MR94 for the stars in common.

Another study at high resolution is by \cite{bes95}, who obtained
echelle spectra of seven K giants in BW, six of which are in common
with our study and one in common with MR94.  The mean difference in
\feh\ (in the sense SRT $-$ Bessell {\it et al.}) is $+0.15 \pm 0.18$.
Combining the results from the \cite{bes95} study and MR84, we find a
mean difference of $+0.12 \pm 0.22$ (s.d.) for the 11 stars in common
with both echelle studies (the scale here is the higher).

We have no stars in common with the study of \cite{cas95}, who obtained
high-resolution spectra of two stars in BW, and who also concluded that
the metallicity scale from Rich (1988) was too high by a few tenths
dex.

Figure 4 shows the correlation between our abundances and those from
both high--resolution studies. There is some indication that the our
metallicities are too high for [Fe/H] $>+0.5$, which is not surprising
since these were generated with a long extrapolation from the
metallicities present in the FFBG libraries.  Since the computed mean
differences are statistically equal to zero, however, we conclude that
the simple FFBG calibration of the \mfe\ index gives a metallicity
scale consistent with the currently available high--resolution studies
in BW, and make no further adjustment to the abundances listed in Table
1.

We can use the stars in common with R88 and GF92 to estimate the true
(external) errors in \feh\ for our sample.  We originally estimated a
mean error of 0.44 dex for our [Fe/H] measurements from propagation of
errors in the \mfe\ index, but we reduced this by a factor of 0.55 to
achieve consistent values of the errors in the three studies.  R88 and
GF92 have seven stars in common, for which
\[
\sigma_{(\rm R88-GF92)} = 0.245 {\rm \ dex},
\] 
where $\sigma$ is the dispersion in $\Delta$\feh\ = [Fe/H]$_{\rm R88} -
$[Fe/H]$_{\rm GF92}$.  From Table 2: 
\begin{eqnarray}
\sigma_{\rm (SRT-R88)}  &=& 0.286 {\rm \ dex\ (48\ stars)},  \nonumber \\ 
\sigma_{\rm (SRT-GF92)} &=& 0.301 {\rm \ dex\ (42\ stars) }, \nonumber 
\end{eqnarray}
after excluding the three discrepant stars noted in Table 3. 
Assuming $\sigma^2_{(\rm SRT-R88)} = \sigma^2_{\rm SRT} + \sigma^2_{\rm
R88}$, where $\sigma_{\rm SRT}$ and $\sigma_{\rm R88}$ are respectively the
errors in the SRT and R88 measurements of [Fe/H], we find:
\begin{eqnarray}
\sigma_{\rm R88}  &=& 0.16 {\rm\ dex }, \nonumber \\
\sigma_{\rm GF92} &=& 0.19 {\rm\ dex }, \nonumber \\
\sigma_{\rm SRT}  &=& 0.24 {\rm\ dex }, \nonumber
\end{eqnarray}

These estimates agree well with the mean errors quoted by R88 and GF92
for their own data, which are 0.18 dex and 0.25 dex respectively.  The
mean of the adjusted errors in Table 1 (reduced by a factor 0.55 from
the original estimates) is 0.24 dex.  It is probably not surprising
that the mean error in our [Fe/H] measurements is slightly higher than
that of R88.  Most of our stars are fainter than those observed by R88,
and the errors reflect the increased effects of crowding as we go
fainter.

\subsection{Reddening and the abundance calibration }
 
We now examine the effects of reddening on our abundance measurements, 
and discuss whether different assumptions about the reddening to BW might 
account for some of the difference between our abundance measurements and 
those of R88 and GF92.  

We consider three questions: \\ 
(i) how do different assumptions about the mean reddening in BW 
affect the derived abundance distribution? \\ 
(ii) is our assumption of a single reddening for the whole BW sample 
reasonable? \\
(iii) what are the effects of patchy reddening for individual stars? 

By examining the dependence of [Fe/H] on temperature in the FFBG
calibration, we find empirically that the effect of small changes in
\evi\ on the derived value of \feh\ varies with temperature, and is
generally larger for hotter stars.  Figure 5 shows the relation between
$\delta{\rm [Fe/H]}$/\evi\ and \teff\ for K giants in our sample.  The
average change in [Fe/H] is 0.15 dex per 0.1 mag in color, with a
maximum value of 0.2.  In Paper I, we estimated the error in the
average reddening for the entire sample to be 0.08 mag.  The maximum
effect of a change of $\pm 0.08$ mag in the adopted reddening would be
a shift of $\pm 0.16$ dex in [Fe/H] (for stars with \teff\ $\sim$
4600\,K), which is somewhat smaller than the uncertainty in an
individual measurement.  (Increasing the reddening would increase
\teff, and since \mfe\ remains constant, the derived value of [Fe/H]
would also increase.)

Could part of the zero--point shift between our data and that of R88
and GF92 be due to differences in the assumed reddening to Baade's
Window?  Both GF92 and MR94 adopted essentially the same reddening as
we did (\evi\ = 0.64 mag; equivalent to E$_{(B-V),0}$ = 0.52 mag).
R88, however, adopted a slightly higher reddening (E$_{(B-V),0}$ = 0.56
mag) and so would be expected to derive a higher abundance for any
given star.  It is difficult to estimate directly the effect of the
higher reddening in the R88 study, since Rich used \jk\ rather than
\vi\ colors to measure temperature, and conversion between the two is
not straightforward.  If we had adopted the higher value of
E$_{(B-V),0}$ = 0.56 mag, our abundances would be higher on average by
0.06 dex.  This is considerably smaller than the measured difference in
the metallicity scales (0.19 dex), from which we conclude that the
differences between our values and those of R88 and G92 does not simply
arise from differences in the adopted reddening to BW.  MR94 discuss the
zero-point of the R88 scale in considerable detail, and conclude that
the R88 scale measures not only the Fe abundance but that of Mg as
well, which may be elevated in bulge stars.  This may also be true of
the G92 scale, which measures essentially the total absorption from all
lines in the blue--yellow portion of the spectrum, including the
strong MgH and Mg{\it b} features.

In Paper I we also showed that there was a {\it possible} variation in
the extinction (on large angular scales) up to $\pm 0.05$ mag across
our BW field.  As discussed above, this should not produce a
significant error in the metallicity scale in different parts of the
sample.  We tested this by computing the mean [Fe/H] in the parts of BW
which Blanco {\it et al.} (1984) identified as having different
extinctions.  In their region A, we find $\langle{\rm [Fe/H]}\rangle =
-0.11 \pm 0.04$, while in the combined regions BC we measure
$\langle{\rm [Fe/H]}\rangle = -0.12 \pm 0.05$. The equality in the two
mean values suggests that there are no significant systematic errors in
our metallicity scale which depend on the positions within BW.

Patchy reddening in the BW field will also affect the abundances we
measure for individual stars.  The general impression from deep
photographs of BW (e.g., Figure 1 of \cite{bla84}) is that some
individual stars may have an extinction which is considerably different
from the average.  Since we cannot estimate the extinction towards
individual stars in our sample, we are unable to determine the
variation in the extinction towards BW on small angular scales.

\section{The line--of--sight to Baade's Window }

The derivation of distances to individual stars in BW is essential to
the interpretation of their proper motions.  The only selection
criteria for inclusion in the BW proper motion sample were that the
stars be relatively uncrowded, and brighter and redder than certain
magnitude limits (\cite{spa92}).

As discussed in Paper I, we expect that our sample will include some
foreground stars in the old disk or thick disk, especially at the
bright end.  This was indeed seen in the data:  the BW stars with $V <
15.5$ have a lower velocity dispersion, a non--zero mean proper motion
in Galactic latitude, and a disk--like anisotropy in velocity
dispersion, consistent with that expected for foreground disk stars at
intermediate distances (Spaenhauer {\it et al.} 1992; Paper I).  In
this section, we derive distances for the stars in our sample, look for
additional signatures of the presence of foreground stars, and provide
several arguments for the plausibility of our approach.

\subsection{What do we expect to see?}
 
What is the likely brightness of bulge and disk giants in Baade's
Window?  Figure 6 shows the predicted {\sl observed}\ V magnitude,
V$_{\rm pred}$, for K\,0 and K\,5 giants (and a K\,0 dwarf) of roughly
solar abundance as a function of distance from the Sun, using absolute
magnitudes M$_V$ from \cite{mih81} and assuming a constant extinction
of A$_V$ = 1.60 mag along the line of sight.  While this is clearly a
simplification, it gives us a rough idea of what we expect to see in
our sample.

We first conclude that our sample is unlikely to contain many disk
dwarfs.  Unless these stars are within 1 kpc of the Sun, they will fall
below the magnitude cutoff of the proper motion sample ($V \sim
17.5$).  Because of Galactic rotation, however, dwarfs within 1 kpc of
the Sun would be expected to show a high proper motion\footnote{The
zero point of the proper motion sample is unknown, so Spaenhauer {\it
et al.\ }(1992) set $\langle\mu_\ell\rangle = 0$, $\langle\mu_b\rangle
= 0$.  Thus if most of the BW sample is near the bulge, disk stars will
show a proper motion about equal in magnitude to the true reflex solar
motion of the galactic center.  At $R = 7.5$ kpc, the Galactic orbital
speed of the Sun, $V_{\rm rot} = 220$ km s$^{-1}$ would produce a
proper motion of 6 mas yr$^{-1}$.} in $l$ relative to the rest of the
sample ($\sim$4.5 mas yr$^{-1}$ at 1 kpc from the Sun, and higher for
closer dwarfs). The mean value for stars with $V < 15.5$ from Paper I
is $\langle \mu_\ell \rangle = 2.75 \pm 0.44$ mas yr$^{-1}$;  while
this is non-zero, as expected for foreground stars, it is considerably
smaller than expected for stars with distances less than 1 kpc.  We
therefore conclude that our sample is unlikely to contain many
foreground dwarfs and that most stars located in front of the bulge
will be {\it giants} in the disk.

If the Galactic Center distance is $R_0 = 7.5$ kpc and the bulge has a
radius of $1-2$ kpc, then bulge giants will have typical apparent
magnitudes of V $\sim 15-17$.  Disk giants closer than 5 kpc will have
$V$ in the range $12-16$.  Disk giants on the far side of the bulge
would have $V > 16-17$.  We do not, however, expect to see many distant
stars in the old disk since the line--of--sight to BW is $\sim 700$ pc
below the disk at 10\,kpc from the Sun;  stars at these distances may
be in the halo or thick disk as we discuss below.

Figure 6 suggests that most stars fainter than $V \sim$ 15.5 are likely
to be in the bulge and most stars brighter than this are likely to be
in the foreground disk, at least for giants of roughly solar abundance.
The large abundance spread in BW provides an extra complication, since
stars of higher abundance will be fainter than stars of lower abundance
at constant color. For $V \sim 15 - 16.5$, therefore, we expect to see
a mix of bulge and disk stars.  We now consider how to identify these
individually.

\subsection{Distance estimates}

We can derive a photometric parallax for each star by comparing its $I$
magnitude to the absolute magnitude $M_{\rm I}$ of a globular cluster
giant of the same (\vi)$_0$ and [Fe/H].  The lefthand panel of Figure 7
shows the mean locus in (\vi)$_0$, M$_{\rm I}$ of the giant branches of
four globular clusters with good--quality color--magnitude diagrams
(CMDs).  These clusters span a range in \feh\ which overlaps the lower
half of metal abundances of the bulge.  They are (from blue to red) NGC
6397 ([Fe/H] $= -1.9$), NGC 1851 ([Fe/H] $= -1.3$) and 47 Tuc ([Fe/H]
$= -0.7$) from \cite{dac90}, plus the metal--rich globular cluster NGC
6553 (\cite{ort90}) for which we adopt [Fe/H] $= -0.2$, \evi\ = 0.77
mag, A$_{\rm I}$ = 1.42 mag.\ and a distance of 4.9 kpc (\cite{bar92};
\cite{bic91}).

For each star in our sample, we interpolated between the cluster CMDs
in Figure 7 to determine the absolute magnitude $M_I$
corresponding to the spectroscopically--determined abundance \feh.
Comparing M$_{\rm I}$ with the dereddened $I_0$ magnitude then gives a
distance modulus for each star.  For stars with [Fe/H]$>-$0.2 (about
half our sample), we had to extrapolate from the NGC 6553 CMD, since
photometry in $VI$ for more metal-rich clusters is not available or
comes from clusters with significant uncertainty in the reddening or
distance modulus.  In the righthand panel of Figure 7, we plotted the
four cluster giant branches (thin solid lines), along with a giant
branch (thick solid line) computed as an extrapolation from NGC 6553.
The adjacent dotted line to that extrapolated giant branch is the
sequence at [Fe/H] $= +0.3$ and an age of 10 Gyr from the Revised Yale
Isochrones (\cite{gre87}).  The nearby points ($\times$) display the
mean giant branch from the photometry of the metal-rich old open
cluster NGC 6791 (\cite{mon94}; \cite{kal95}) which has been plotted
using the values of reddening and distance modulus advocated for this
cluster by \cite{tri95}, who derive [Fe/H] $= +0.36 \pm 0.12$ (we
prefered not to use the cluster photometry directly in our computation
of distances because the available data extends only $\sim 1$ mag above
the cluster horizontal branch).  The giant branches of NGC 6791, from
the RYI, and as extrapolated in our computation agree on average to a
few hundredths mag in $M_I$, suggesting that up to this metallicity our
computation of distances does not contain significant systematic
errors.  Beyond [Fe/H] $= +0.3$, of course, the extrapolation is
uncertain and any distance estimate must be regarded with caution.

\subsection{The Bulge and the Disk}

We now show that our sample contains subgroups which may be identified
with giant--branch and clump stars in the bulge, giants in the
foreground disk, and a few stars from the halo beyond the bulge.  Our
aim here is to identify likely bulge stars in a way which is
independent of their kinematics.  In the solar neighborhood, the two
main stellar populations (the disk and halo) can be distinguished
either by abundance or by kinematics, but it is not possible to
separate the bulge and disk populations in Baade's Window (BW) in the
same way.  The mean abundance of K giants in BW is similar to that of
of the local disk (MR94), so we cannot reliably distinguish between
bulge and disk stars on the basis of abundance.  Radial velocity
measurements by themselves are not very helpful either --- bulge stars
have a high velocity dispersion ($\sim$ 110 km s$^{-1}$), while the
velocity dispersion of the old disk along the line--of--sight to BW
increases from 40 km s$^{-1}$ near the Sun to $\sim 90$ km s$^{-1}$
near the Galactic center (\cite{lew89}).

We do not intend in the following discussion to imply that every star
can be unambiguously placed into distinct population bins;  rather we
argue that the combination of metallicities and distances which we have
derived allows us to identify the principal subgroups of our sample in
a way which is consistent with current information about the structure
and kinematics of the Milky Way.  The numerical values we derive below
for metallicities and velocity dispersions are summarized in Table 4.

\subsubsection{Clump/RHB stars in the bulge}

We first note that the photometric parallax method we describe above
may be systematically in error for the clump or red horizontal branch (RHB)
giants which are numerous in BW (\cite{ter88}; \cite{pac94}, hereafter
P94; \cite{tie95}).  Clump giants appear on CMDs of the bulge at
somewhat bluer colors than the center of the giant branch (equivalently,
at brighter magnitudes at each color), so distances
estimated from their photometric parallax may be systematically low.
We used the location of the RHB/clump in 47 Tuc and NGC 6553 to
estimate the region of the CMD in BW where most of the clump/RHB stars
lie.  To allow for the distance spread within the bulge, we took a
range of M$_{\rm I}$(7.5) $-$0.8 to $+$0.1 for the bulge clump/RHB
stars, where M$_{\rm I}$(7.5) is the magnitude each star would have at
a distance of 7.5\,kpc.  Note that these stars would be either clump
stars or first--ascent giants in the vicinity of the clump.  It is also
possible that this narrow magnitude range contains a few disk giants.

On this basis, we estimate that 196/322, or 61\% of our sample, are
clump/RHB giants.  Most of these stars have $V$ between 16.5 and 17.5,
which is the location of the clump on our CMDs (Figure 1 of Paper 1).
This appears at first to be a very high fraction, but P94 also note
that red clump stars outnumber red giant--branch stars in their
photometric samples of several BW fields.  P94 use slightly different
criteria to distinguish clump and giant--branch stars, but the results
are very similar: applying their selection criteria (1) and (2) to our
sample gives 222/322, or 69\%, clump stars.  We measure a weighted
velocity dispersion (using the prescription in Armandroff \& DaCosta
1986) of $111 \pm 13$ km s$^{-1}$ (see Table 4), which agrees well with
that measured for other stellar samples in Baade's Window
(\cite{ric90}; \cite{sha90}), and SiO masers in the same region of the
bulge (\cite{izu95}). Thus the group we identify as clump/RHB stars
appears kinematically similar to other bulge populations.
W\subsubsection{Giants in the Bulge}

Since the Galactic bulge is spatially extended, stars in the bulge will
span a range in distance.  We take $\sim 1$ kpc as the characteristic
radius of the Galactic nuclear bulge, based on the distribution of M
giants (\cite{bla89}), IRAS point sources (\cite{hab85}; \cite{har88})
and the infrared surface brightness of the bulge (\cite{ken92};
\cite{wei94}).  Thus we would expect our distance measurements for
bulge giants to span a range of at least $R = 7.5 \pm 1$\,kpc.  There
will, however, be an extra spread due to distance errors.  The mean
error in [Fe/H] for an individual star in our sample is $\sim 0.24$
dex, which corresponds to $\sim \pm0.3$ mag in the distance modulus.  A
realistic distance spread (taking into account both the depth of the
bulge and the uncertainty in determining [Fe/H] from the measured \mfe)
is therefore $7.5 \pm 2.5$ kpc, so that stars with $R$ in the range
5--10 kpc are likely to be bulge giants.  Of the 126 stars in Table 1
which are not clump stars, 72 fall into this class.  They have a mean
distance of 7.0 kpc, and a median distance of 7.1 kpc.  The velocity
dispersion for this group is $95 \pm 17$ km s$^{-1}$, slightly lower
than but equal to within the errors to the value for the clump/RGB
subsample.

\subsubsection{Foreground disk giants}  

We next identify as stars in the foreground disk as those which have an
estimated distance R $<$ 5\,kpc, and which are not clump/RHB stars.
There are 39 such stars (12\% of our sample), with a mean distance of
3.3 kpc from the sun.

The radial velocity dispersion of the disk giants, $82 \pm 21$ km
s$^{-1}$, is lower than that of the clump/RHB stars but higher than
Lewis \& Freeman's (1989) predicted value of $\approx 72$ km s$^{-1}$
for the velocity dispersion of old disk stars at the mean distance of
our disk sample;  it is possible that the higher velocity dispersion
indicates that there are some stars in the metal-rich end of the thick
disk in our sample at low distances.  Sharples {\it et al.} (1990)
noted that the bright M giants toward BW had a lower velocity
dispersion than the bulk of the sample, and attributed these stars to
the foreground disk.  They found $\sigma(v_r) = 71^{+20}_{-11}$ km
s$^{-1}$ for the bright stars, consistent within the errors to our
value for the disk subsample in this survey.

\subsubsection{Halo stars}

The remaining 15 stars in our sample (5\%) have R $>$ 10 kpc, and so
appear to lie beyond the bulge.  As mentioned earlier, we would not
expect to see disk stars at such large distances because the
line--of--sight is so far from the plane.  These distant stars have a
low mean abundance and a high velocity dispersion, suggesting that many
of them belong to the Galactic halo. It is also possible that some are
bulge stars with much higher reddening than average (and therefore
whose distance is over--estimated), but this would require that some
small patches in BW have $E_{V-I}$ which is up to $\sim 0.5$ mag
higher than the average value.

\subsection{The Distance Distribution in Baade's window} 

Table 1 lists the estimated distance for each star, and a
classification as bulge clump/RHB (C), bulge giant branch (B), disk
giant branch (D) or halo giant (H).  In the discussion which follows,
the term `bulge stars' refers to the combined B+C sample which contains
268 stars, i.e.\ 83\% of the stars with metallicity estimates.  Table 4
summarizes the basic properties of the C, B, D and H subsamples.

Figure 8 shows the distribution of estimated distances for the entire
sample of 322 stars (dashed line), as well as the distribution with
with the clump/RHB stars excluded (solid line).  The latter
distribution is more representative of the distance distribution in the
BW sample because of the systematic uncertainty in the distance
estimates for clump/RHB stars.  This B+D+H subsample has a mean
distance of 6.6\,kpc and a median distance of 6.4 kpc.  This is
slightly lower than our adopted Galactic Center distance of 7.5 kpc
both because of the inclusion of foreground disk giants and because the
magnitude cutoff of the proper motion sample means that we may not see
the most distant of the metal--rich bulge stars.

We see a slight excess of disk giants in the R = 2$-$3\,kpc bin in
Figure 8.  A similar feature was noted by by P94 in their study of CMDs
in BW (cf.  \cite{ter88}), and attributed to spiral arm structure in
the galactic disk.

As mentioned above, our distance estimates for the clump/RHB stars will
be systematically low because these stars lie $\sim 0.5 - 0.7$ mag
above the giant branch stars of similar color in the CMD (\cite{ter88};
P94; \cite{ng95}).  The mean distances for the B and C subsamples, 7.1
and 5.7 kpc, respectively, confirm this. The difference in distance
modulus for the mean of two samples is 0.48 mag, consistent from the
difference expected from the CMDs in the bulge.  In Figure 8 we present
an alternate distance distribution for our sample, which was
constructed by arbitrarily increasing the distances of the C group by a
factor of 1.25 to bring the mean distance equal to that of the B
subsample.  The resulting distribution is shown in Figure 9.  The mean
and median distances are, respectively, 7.6 kpc and 6.9 kpc.

Figure 10 shows the observed contribution of each subgroup to the BW
sample as a function of $V$ magnitude.  Foreground disk giants
contribute roughly 50\% of the sample at $V = 15$, 20\% at $V = 16$ and
are almost absent by $V = 17$.  Bulge clump stars occupy a narrow range
between $V= 16.5$ and 17.5.

\section{The abundance distribution of K giants in Baade's Window }

\subsection{The \feh\ distribution }

Figure 11 shows a histogram of \feh\ for the 268 bulge stars with
abundance measurements (classes B and C above).  The weighted mean
abundance for this group is $\langle{\rm [Fe/H]}\rangle = -0.11 \pm
0.04$, with a dispersion of $\sigma({\rm [Fe/H]}) = 0.46 \pm 0.06$ dex;
the median abundance is $-0.03$ dex. Over half the stars lie in the
range \feh\ = $-0.4$ to $+0.3$.

A potential problem with the derivation of abundances from our
low--resolution spectra is that the FFBG formulation is only valid for
$-0.8 \leq {\rm [Fe/H]} \leq +0.5$.  Above \feh\ $+0.5$ the abundance
estimates are extrapolations, while below \feh\ $-0.8$ the iron lines
measured by the \mfe\ index become too weak to yield reliable results.
About 18\% of our sample have derived [Fe/H] values which lie outside
this range of validity, but estimates of the mean abundance in our
sample are unlikely to be seriously affected by their inclusion.

The mean abundance for our sample of 268 bulge K giants is somewhat
higher than the value of $\langle{\rm [Fe/H]}\rangle = -0.25$ derived
by MR94 from a recalibration of the 88 giants observed by R88.  If we
compare our \feh\ with the recalibrated \feh\ for the R88 stars in
common with our sample using equation (6) of MR94, we find
\[
\langle{\rm [Fe/H]_{SRT}} - {\rm [Fe/H]_{R88'}}\rangle
 = 0.14 \pm 0.23~{\rm (s.d.)},
\]
where the recalibrated Rich (1988) scale as denoted ${\rm R88'}$.
Almost all of this difference is from stars with [Fe/H] $> +0.3$, for
which in comparison to the echelle metallicities in BW (see above) our
calibration is possibly too high.  Taking stars with ${\rm [Fe/H]} \leq
+0.3$ in our solution only, we find
\[
\langle{\rm [Fe/H]_{SRT}} - {\rm [Fe/H]_{R88'}}\rangle
 = 0.02 \pm 0.15~{\rm (s.d.)}
\]
Thus although our measured values of [Fe/H] agree well with those
measured by MR94 for the few stars we have in common, it appears that
our \feh\ scale could be systematically higher by $\sim$ 0.15 dex than
the R88 scale as recalibrated by MR94, and that the difference is
primarily because we may be overestimating the metallicities at the top
end.  This difference is about equal to that found by comparison to the
results of high-resolution studies in BW (Sec. 2.4, above).  However,
we do not think it appropriate to recalibrate our \feh\ measurements at
this stage. The MR94 recalibration of the R88 scale is based on only 11
stars in common, and a larger sample of good--quality echelle spectra
is clearly needed to set the low--resolution abundance measurements on
a firm basis. For now, we simply accept that the zero point of our
[Fe/H] scale is uncertain by 0.10 to 0.15 dex, and bear this in mind in
the discussion which follows.

\subsection{Abundance measurements from Mg and CN indices}

As well as deriving [Fe/H] from the \mfe\ index, we can use the Mg$_2$
and CN indices as alternative abundance indicators.  If the ratios
[Mg/Fe] and [(C+N)/Fe] have the same values in BW stars as in the Sun
(as the FFBG relations assume), then we should derive the same mean
abundance for [Mg/H] and [CN/H] as for [Fe/H] in the sample as a
whole.  There are, however, reasons to expect {\it non--solar}
abundance ratios in many of our BW giants.  Old stellar populations in
the Galaxy generally have ${\rm [Mg/Fe]} > 0$, at least for abundances
well below solar (e.g., \cite{lam87}; \cite{whe89}).  In BW, there is
evidence from both low--resolution spectra (Paper I) and echelle
studies (MR94) for an enhancement of [Mg/Fe] in bulge giants. The Mg
and Fe line strengths in the integrated light of elliptical galaxies
and spiral bulges also show evidence for a Mg overabundance
(\cite{wor92}).

Once again, we use the relations tabulated by FFBG to convert our
measured line--strength indices to abundances.  By doing this, we are
implicitly assuming that the BW giants have the same values of {\it
gravity} as a function of temperature as the calibrating stars (mostly
solar-neighborhood giants) in FFBG;  using the G93 formulations, with
different assumptions about the gravities of the BW giants, gives
somewhat different results which we will discuss below (Sec. 4.5).  We
calculated abundances from the Mg$_2$ and CN indices\footnote{The
abundances from the Mg$_2$ and CN indices will be denoted [Mg/H] and
[CN/H], and the relative abundances with respect to the metallicities
from \mfe\ will be denoted [Mg/Fe] and [CN/Fe].  The reader is reminded
that we are not deriving abundance ratios in any true sense.  We are
comparing metallicity {\it indicators} from different features;  these
probably should have been called M($\langle{\rm Fe}\rangle$),
M(Mg$_2$), and M(CN).  Our notation for the abundance ratio is a
convenient shorthand:  e.g., [CN/Fe] $=$ M(CN) $-$ M($\langle{\rm
Fe}\rangle$).} for all stars where these indices were measurable, but
chose to analyze only the set of 195 stars for which the errors in all
three abundance measurements were less than 0.5 dex.  Even for these
stars, the errors in the individual abundance ratios [Mg/Fe] and
[CN/Fe] are quite high (typically 0.4 to 0.5 dex), but the sample as a
whole shows some interesting trends which are worth discussing.  Table
5 summarizes the mean abundance ratios for several subgroups.

Figure 12 shows the FFBG calibration of abundance for Mg$_2$ and CN as a
function of \teff\ (compare to Figure 5).  The Mg$_2$ index is strongly
temperature--sensitive, while the CN index is almost independent of
temperature for K giants.  It is clear from Figure 12 that the BW stars
in our sample do not have the same abundances on the three scales:  the
mean [Mg/H] is near solar, as was [Fe/H], but the mean [CN/H] is
considerably lower, with $\langle{\rm [CN/H]\rangle} \sim -0.3$.

\subsection{[Mg/H]}

Figure 13 (lower) shows a histogram of [Mg/H] for the 159 bulge stars
in the (B + C) groups which have reasonably small errors from all three
indices.  The distribution is similar to that of [Fe/H] in
Figure 5.  Both the [Mg/H] and [Fe/H] distributions have a dispersion
of 0.50 dex.  The mean value of [Mg/Fe] for this group is $0.08 \pm
0.03$ dex.

If the bulge sample is divided at [Fe/H] = 0.0, the metal--poor stars
have a higher mean value of [Mg/Fe] than the metal--rich stars (see
Table 5).  Figure 14 (lower) shows the relation between [Mg/Fe] and
[Fe/H] for individual bulge stars.  Stars with [Fe/H]$>$0 show no
strong Mg excess, and the behavior of the bulge stars as a whole
appears similar to that of stars in the old disk and halo (e.g.,
Lambert 1987).

The Mg overabundances which we derive are modest, and lower than our
estimate of $\sim +0.3$ dex in Paper I based on plots of line--strength
indices and some simple models.  They are also lower than the typical
values measured by MR94.  For the 5 stars in common (our measurement of
[Mg/H] for star 3$-$152 is unreliable), we find $\langle{\rm
[Mg/Fe]}\rangle = +0.03 \pm 0.06$, significantly lower than the value
of $+0.31 \pm 0.05$ measured by MR94.  For 49 stars in the range [Fe/H]
$-0.5$ to $0.0$, we find $\langle{\rm [Mg/Fe]}\rangle = +0.11 \pm
0.06$, compared to typical values of $+0.4$ dex measured by MR for
stars in the same [Fe/H] range.

\subsection{[CN/H]}

The behavior of [CN/H] shows several puzzling features.  At first
sight, CN is potentially the most robust of the line--strength indices
we have measured, since (Figure 12, top panel), it is almost
independent of temperature for K giants and thus is unaffected by small
errors in the assumed temperature or reddening.  Detailed examination
of the CN strengths in field and cluster K giants, however, shows that
CN is affected by other factors, most likely core/surface mixing (G93
and references therein).

The distribution of [CN/H] in Figure 13 is narrower than that for
[Mg/H] and [Fe/H] (see Figure 2), and peaks at a significantly lower
abundance.  This is surprising, since mixing would be expected to
increase the observed abundance of CNO elements and differences in the
amount of mixing along the giant branch might be expected to broaden
the observed [CN/H] distribution so that it was wider than that of
[Fe/H].

For the metal--poor and metal--rich subgroups of our bulge stars, the
average [CN/Fe] values are very different.  The metal--poor stars
(Table 5) have a mean [CN/Fe] ratio close to solar, while the
metal--rich stars have $\langle{\rm [CN/Fe]}\rangle = -0.47 \pm 0.04$.
These mean values are plotted with the individual determinations of
[CN/Fe] on the top panel of Figure 14.

R88 also noted that CN was weak relative to Fe in many of the BW stars
in his sample (see also \cite{jan77}). He thought that surface gravity
effects were unlikely to be the cause, since the Mg$_2$ index (which,
like CN, is gravity--sensitive) showed no tendency for CN--weak stars
to have the higher gravities which would be needed to depress the CN
band.

We considered the possibility that [CN/H] reflects the `true' abundance
distribution of BW giants (i.e., a narrow range around a mean abundance
of $-0.3$\,dex, or 0.6 times solar), and that the derived [Mg/H] and
[Fe/H] values are distorted by both patchy reddening in BW (which would
broaden the distribution) and an assumed mean reddening which is too
high.  This appears unlikely.  Comparing the observed \vi\ color and
H$\beta$ line index (as described in Paper I) for stars with low and
high [CN/Fe] shows no evidence for a difference in reddening between
the two groups.

\subsection{Exploring the effects of gravity}

As mentioned earlier, a fixed relation between temperature and surface
gravity is implicit in the FFBG calibration which we use to convert our
line--strength measurements to abundances.  Since bulge K giants may
not necessarily follow the same $\log g$--\teff\ relation as those in
the solar neighborhood, we explored the effects of this assumption by
recomputing [Fe/H], [Mg/Fe], and [CN/Fe] from the G93 formalism, which
explicitly includes both metallicity and gravity in its ``fitting
functions'' for each line--strength index.

We first estimated $\log g$ and \vk\ for each star (the latter is the
G93 temperature indicator).  For gravity, we made an initial estimate
by assuming that each star was at a distance of $R = 7.5$ kpc, and
derived $\log g$ from
\[
\log(g/g_\odot) = \log(M/M_\odot)
                - \log(L/L_\odot)
                + 4 \log(T/T_\odot),
\]
where $M$, $L$, and $T$ are, respectively, the stellar mass, luminosity
and temperature.  We assumed $M = 1.1\,M_\odot$, a value appropriate
for moderately old stars (this will be lower than the values implicit
in the FFBG approach), and derived $T$ from the \vi\ color as above.
The luminosity was derived from the dereddened $I$ magnitude with
bolometric corrections appropriate for bulge stars (\cite{tie95}), and
the \vk\ color was derived from the dereddened \vi\ using a polynomial
relation derived from a small set of stars in BW which have both colors
(\cite{tie95}; Paper I).  We then estimated [Fe/H] for each star using
the mean of the values from the G93 fitting functions for the Fe\,5270
and Fe\,5335 indices.

The \feh\ estimates derived in this way agree well with our earlier
determinations using the FFBG relations.  The G93 values are tightly
correlated with the FFBG values (as expected, since they are derived
from the same indices), but $\approx$0.05 dex higher.  With these new
\feh\ values we computed the distance to each star as described
earlier, which led to a new estimate of $L$ and hence $\log g$.  These
were then inserted into the fitting functions for the Mg$_2$ and CN
indices to derive estimates of [Mg/H] and [CN/H].  These were
differenced with [Fe/H] to provide [Mg/Fe] and [CN/Fe].  The earlier
remarks on the large errors in these quantities apply to this
computation as well.

Finally, we divided the sample into disk, bulge, and halo groups as
described earlier in Sec.\ 3 and computed mean values of [Mg/Fe] and
[CN/Fe] for each group.  These are listed in Table 5 together with the
earlier values from the FFBG approach.

In Figure 15 we display the derived mean values of (lower panel)
$\langle{\rm [Mg/Fe]} \rangle$ and (upper panel) $\langle{\rm [CN/Fe]}
\rangle$ as a function of \feh\ for the stars in the bulge (B + C)
sample.  These were computed by sorting the stars by [Fe/H], and
dividing the stars into four bins of equal numbers.
In both panels of Figure 15, the mean abundance ratios
as computed in the FFBG approach are shown as open points with
error bars (these are errors in the mean).  The filled points
connected with the dashed line are the mean abundance ratios
from the G93 formalism.  In the lower panel, we also show (solid
line) the trend in the relative abundances of the alpha-capture
elements in the local disk and halo (e.g., \cite{whe89}).  The
plusses show the derived [Mg/Fe] ratios for individual K giants
from MR94.

For CN, the two approaches give similar results. In both cases, we find
$\langle{\rm [CN/Fe]} \rangle \approx 0$ for the most metal--poor stars
in our sample, and $\langle{\rm [CN/Fe]} \rangle \approx -0.4$ above
[Fe/H] $= 0.0$.  Thus the low abundances derived from the CN index for
metal--rich stars in our sample do not appear to be related to
surface--gravity effects.

For Mg, the G93 formalism gives [Mg/Fe] values up to 0.3 dex higher
than the FFBG approach.  As G93 showed, the Mg$_2$ index is strongly
affected by surface gravity while the effect on the strengths of the Fe
lines is negligible.  In the FFBG approach, the metallicity calibration
of the Mg$_2$ index was made under assumption that [Fe/H] $\equiv$
[Mg/H], using adopted metallicities from K giants in the solar
neighborhood.  If these stars are significantly younger than K giants
in the bulge, they would have higher surface gravities (and therefore
stronger Mg absorption lines) at a given luminosity.  For the G93
calculation of [Mg/Fe], we assumed a lower gravity (appropriate for
relatively old stars in the bulge), which means that we would derive
higher [Mg/Fe] to match the observed Mg$_2$ values in our sample (the
Mg$_2$ index increases with metallicity and with gravity).

That the K giants in the solar neighborhood probably have higher
gravities than bulge stars may be seen directly in Figure 3 of G93,
which plots $\log g$ versus temperature for their calibration stars.
At colors typical of our sample, the gravities of stars in old globular
clusters like M\,3 and M\,92 are $\sim 1.5$ dex lower than those for
stars in metal--rich disk clusters like M\,67 and M\,71.  We derive
higher values of [Mg/Fe] (and better agreement with the results of
MR94) if we use the G93 fitting functions and assume that the BW giants
are substantially older than solar neighborhood K giants, i.e., their
gravities are lower than for the K giants in the FFBG formulation.  It
is clear that the values of [Mg/Fe] derived from low--resolution
spectra are strongly dependent on assumptions about the surface
gravities of bulge K giants, which cannot be measured directly from the
low--resolution data (see also Figure 11 of Paper I).

\section{Summary and Discussion}

In this paper, the second in our series on the kinematics and
abundances of K giants in the Baade's Window field of the Galactic
bulge, we have used Fe line strengths to measure [Fe/H] for 322 stars.
The derived abundance scale is close to that from two recent studies of
bulge stars at high resolution.  In preparation for our upcoming
analysis of the proper motions and radial velocities of our sample, we
derived photometric parallaxes and showed that the line--of-sight
towards BW contains subgroups which we can identify with the foreground
disk, clump and red giant stars in the bulge, and distant halo stars.
About 80\% of our sample can be classified as bulge giant--branch or
clump/RHB stars.

There has been considerable discussion in the literature about the mean
metallicity in the bulge.  Several earlier studies, based on line
strengths in K giants (\cite{whi83}; \cite{ric88}), Washington
photometry (\cite{gei92}) and the photometric and spectroscopic
properties of M giants (\cite{fro87}; \cite{fro90}; \cite{ter90};
\cite{sha90}; \cite{ter91}) concluded that the mean abundance of the
stars in BW was $\sim +0.2$.  This was revised sharply downward to
$\langle{\rm [Fe/H]}\rangle \approx -0.25$ by MR94 based on their
high--resolution studies of 11 K giants in BW, and a recent study of
the infrared color--magnitude diagram in BW (\cite{tie95}) yields much
the same result.  MR94 give a lengthy discussion of why their estimated
metallicities are lower, concluding that the R88 scale measured [(Mg
$+$ Fe)/H], with [Mg/Fe] $> 0$ in the bulge;  they also found elevated
relative abundances of Ti in their sample, and speculated that if this
were the case for the bulge M giants, the average abundances found for
these stars would be higher than those from Fe.

In this paper, we find $\langle{\rm [Fe/H]} \rangle = -0.11 \pm 0.03$
for our combined sample of bulge giants and clump/RHB stars.  Our
[Fe/H] measurements agree well with those measured at high resolution
for the few stars in common; but appear to be $\sim$0.15 dex higher
than the R88 scale as recalibrated by MR94, though much closer
for stars with [Fe/H] $\leq +0.3$.  Because the recalibration
is based on a small number of stars we have chosen not to
apply any correction to our measured \feh\ values at this stage.

We then explored whether the metallicity scales derived from the CN and
Mg$_2$ indices are the same as those from the Fe features.  There is
evidence that the metal--poor half of our sample has [Mg/Fe] $> 0$, but
in general the Mg excess is less than that found by MR94.  The FFBG
formalism which we used to derive [Mg/Fe] from our low--resolution
spectra assumes that bulge K giants have the same surface gravity as
giants of similar abundance in the local disk.  This is unlikely to be
the case if most bulge giants are relatively old (see also Paper I), so
it is possible that we have underestimated [Mg/Fe] for many of our
stars.

The behavior of CN is puzzling --- many of our stars, particularly the
the most metal--rich ones, appear to be CN--weak compared to
local giants. G93 discussed CN
extensively, and argued that the primary controller for the CN index
was mixing at the base of the giant branch.  They found a strong rise
in CN line strength beginning at about \vk\ = 2.1, and essentially
complete by \vk\ = 2.2.  These colors correspond to the {\it bluest}
stars in our sample, so that if the bulge stars have the same
combination of metallicities and/or mixing as the K giants in the Lick
program, we would expect to see similarly strong CN features.  We find,
however, near solar [CN/Fe] for the metal--poor half of the bulge
sample but weak CN for stars with [Fe/H] $> 0$.

The weak CN lines suggest that bulge stars may differ in some respects
from the stars which make up ellipticals and the bulges of other
spirals.  For Mg, Worthey {\it et al.} (1992) have shown that the
nuclear spectra of most ellipticals have higher Mg$_2$ indices compared
to iron, which cannot be explained with models of the integrated light
which have [Mg/Fe] $= 0$.  They suggest that Mg is enhanced in many
external systems, and that the Mg overabundance increases with galaxy
luminosity.  In Galactic bulge giants, CN appears to be much weaker
than in the bulge of M\,31, which has a high CN index (Worthey 1994).
The behavior of CN in external galaxies is not well understood.   For
example the Worthey models at the highest metallicity ([Fe/H] $= +0.3$)
always produce much stronger indices than observed in galaxies,
suggesting that the mean abundance is less than in the models.  The
exception to this is CN, where the observed line strength can be higher
than in any model, suggesting that it is not just the overall
metallicity which varies.

\acknowledgments

We wish to thank Gary DaCosta, Ken Freeman, Jeremy Mould, John Norris
and Ruth Peterson for advice and helpful discussions.  RMR received
support from NASA Grant NAGW 2479.  DT received support from the
National Science Foundation through grants AST-9157038 and INT-9215844.
We are grateful to an anonymous referee for several valuable
comments on the original version of this paper.

\clearpage

\begin{figure}
\caption{Relation between the reddening--corrected color (\vi)$_0$\ and
effective temperature \teff, from Bessell (1979).} 
\end{figure}

\begin{figure}
\caption{Conversion between \teff, \mfe\ and \feh, adapted from Faber 
\etal(1985). The 322 individual data points with \teff\ in the range
3900--5160\,K are also shown. } 
\end{figure}

\begin{figure}
\caption{Comparison between \feh\ measurements from this study and
those measured by Rich (1988; lower panel) and Geisler \& Friel (1992;
upper panel) for stars in common.  The dotted lines show constant
offsets of $+0.2$ and $0.15$ dex respectively.}
\end{figure}

\begin{figure}
\caption{Comparison between \feh\ measurements from this study, and
measurements from high--resolution spectra studies by (filled circles) 
McWilliam \& Rich (1994) and (open circles) Bessell \etal(1995).  
The dashed line denotes identity. } 
\end{figure}

\begin{figure}
\caption{Effect of changes in \evi\ on the measured \feh\ value for
stars of different temperatures.}
\end{figure}

\begin{figure}
\caption{Predicted observed $V$ magnitude for K0 and K5 giants (solid
lines) and a K0 dwarf (dashed line) at various distances along the
line--of--sight in BW.}
\end{figure}

\begin{figure} 
\caption{Adopted absolute magnitude sequences.  Shown in the left panel
are isochrones for the four globular clusters described in the text
(solid lines).  The two dotted lines show the region in which we have
assumed that the bulge clump stars lie. Open circles represent stars in
our sample, assuming that all the stars are distance of 7.5 kpc.  The
right panel repeats the globular cluster sequences (thin solid lines)
and the clump region (horizontal dotted lines).  The thick solid line
shows the giant branch for [Fe/H] $= +0.3$ as extrapolated from the
other giant branches as described in the next.  The adjacent dotted
giant branch is for the Revised Yale Isochrone (Green {\it et al.}
1987) at [Fe/H] $= +0.3$ and an age of 10 Gyr.  The points ($\times$)
represent the observed giant branch for the old metal-rich open cluster
NGC 6791 (Kaluzny \& Rucinski 1995; Montgomery {\it et al.} 1995)
shifted to the reddening and distance modulus advocated by Trippico
{\it et al.} (1995), who derive [Fe/H] $= +0.36 \pm 0.12$ for
NGC 6791.}
\end{figure}

\begin{figure}
\caption{Distribution of estimated distances for the entire sample
(dashed line), and for stars on the giant branch only, with clump stars
excluded (solid line). The bar shows the typical error in an individual
distance measurement.}
\end{figure}

\begin{figure}
\caption{Alternative distance distribution for all stars in the
survey, constructed by increasing the distances of the clump/RHB
stars in Figure 7 by a factor of 1.25.}
\end{figure}

\begin{figure}
\caption{Relative contribution of bulge, disk and halo stars to the
stellar population in BW, as a function of observed apparent $V$ magnitude.}
\end{figure} 

\begin{figure}
\caption{Histogram of [Fe/H] for the 268 stars which are members of the
bulge (i.e. bulge giant--branch and clump stars).}
\end{figure}

\begin{figure}
\caption{Conversion between \teff, line strength and abundance for the
Mg$_2$ and CN indices, adapted from Faber \etal(1985). As in Figure 2,
the 322 individual data points with \teff in the range 3900--5160 K
are also shown.}
\end{figure}

\begin{figure}
\caption{Histograms of [Mg/H] (lower) and [CN/H] (upper) for bulge
stars in our sample.}
\end{figure}

\begin{figure}
\caption{Plots of the abundance ratios [Mg/Fe] (lower) and [CN/Fe]
(upper) from the FFBG calibaration as a function of [Fe/H] for
individual stars (shown as crosses).  The large open squares show the
mean values for stars with [Fe/H] $<$0 and $\geq$ 0, while the solid
line shows the relation between the Mg and Fe abundances for local
stars in the disk and halo (Lambert 1987).}
\end{figure}

\begin{figure}
\caption{Basic trends in abundances from the Mg$_2$ index (lower panel)
and CN index (upper panel).  The bulge (B+C) stars have been split into
four equal bins in \feh.  The mean values from the simpler FFBG
calibration (Faber {\it et al.\ }1985) calibration are shown by open
circles, while the filled circles show the mean [CN/Fe] and [Mg/Fe]
abundances derived from the G93 (Gorgas {\it et al.}\ 1993) formalism.
In the lower panel, plus signs denote individual [Mg/Fe] values from
the study by McWilliam \& Rich (1994) and (as in Figure 14) the solid
line shows the [Mg/Fe] trends in the local disk and halo (Lambert
1987).}
\end{figure}

\clearpage
{\sc TABLE} 1.  Measurements of [Fe/H] for Baade's Window K--giants. 

{\sc TABLE} 2.  Comparison of [Fe/H] measurements for stars in common with
                other studies. 

{\sc TABLE} 3.  Cross--identification with stars observed by Geisler \& Friel
                (1992). Asterisks mark the three stars with highly discrepant 
                [Fe/H] meaurements. 

{\sc TABLE} 4.  Mean abundance, and radial velocity dispersion, for subsamples 
                of the BW population.  

{\sc TABLE} 5.  Mean abundance ratios for BW subsamples. 

\end{document}